\input{aipcheck.tex}

\def\selectedoptions{final}

\documentclass[
   \selectedoptions
  ]
  {aipproc}

\def\selectedlayoutstyle{6x9}

\layoutstyle\selectedlayoutstyle

\SetInternalRegister\hbadness{8000} 

\newcommand\doingARLO[2][]{%
  \ifx\mmref\undefined #1\else #2\fi
}

\begin{document}


\title 
{Pion Corrections in Gribov's Approach to the Dyson-Schwinger Equation}

\author{Carlo Ewerz}{
  address={Universit\`a di Milano and INFN, Via Celoria 16 , I-20133 Milano, Italy},
}

\begin{abstract}

Chiral symmetry breaking in QCD leads to the 
emergence of pions as Goldstone bosons. Their existence 
in turn affects the Green functions of the theory. 
Here we study the effect of pion corrections on the light quark's Green 
function in Feynman gauge using 
the framework of Gribov's approach to the 
Dyson-Schwinger equation. 

\end{abstract}

\date{28.\ November 2004}

\maketitle

\section{Gribov's Approach}

The system of Dyson-Schwinger equations (DSEs) for the 
Green functions is one of the few tools we have for 
investigating the nonperturbative structure of QCD. In particular, 
one would like to use the DSEs to understand the mechanism 
giving rise to chiral symmetry breaking ($\chi$SB) and confinement. 
The difficulty in doing so lies in the fact that the DSEs form a  
tower of coupled integral equations. Their treatment 
hence requires truncations or approximations 
which are in general difficult to control. 

Some time ago Gribov suggested a new approach to the 
DSE for light quarks \cite{Gribov:1998kb}. 
His approach is designed to systematically collect the most important 
contributions to this integral equation originating from the infrared (IR) 
region in which the dynamics leading to $\chi$SB 
and confinement is expected to take place. The largest contributions to 
the integrals in the DSE from the IR can be most conveniently 
identified in Feynman gauge. In that gauge the gluon propagator has the 
form $D_{\mu\nu}(k) = - \alpha_s(k)g_{\mu\nu}/k^2$, which 
can be understood as a definition of a nonperturbative coupling 
$\alpha_s(k)$ at small momenta. The applicability of Gribov's approach 
requires only mild assumptions on the behavior of $\alpha_s(k)$, namely 
that is does not diverge at $k=0$ and does not vary too rapidly with $k$. 
One then takes the second derivative of the inverse Green function, 
$\partial^\mu \partial_\mu G^{-1}(q)$ with 
$\partial_\mu= \partial/\partial q^\mu$. 
Using the DSE one finds that the most singular contribution 
from the IR comes from differentiating twice the gluon Green function 
because $\partial^2(1/q^2) \sim \delta(q^2)$. 
With the help of Ward identities one then 
obtains a differential equation for the light quark's Green function, 
see eq.\ (\ref{gribovgl}) below without the last term. 
It can also be shown to reproduce the correct 
renormalization group equation in the ultraviolet (UV) region. 
Less IR-singular terms can be computed systematically in this 
approach as subleading corrections. 
In this sense Gribov's approach is an approximation rather 
than a mere truncation scheme for the quark's DSE. 

One can then use Gribov's equation for a detailed investigation 
of the light quark's Green function in Feynman gauge. 
According to \cite{Gribov:1998kb} it should in particular be 
possible to use this approach in order to study the analytic properties 
of the Green function and to relate them to a picture in which 
confinement is caused by the phenomenon of supercritical charges. 
For a recent review of the ideas underlying Gribov's picture 
of confinement we refer the reader to \cite{Dokshitzer:2004ie}. 

\section{Pions from Chiral Symmetry Breaking}

As was discussed at a previous edition of this conference
\cite{Ewerz:2003wx}, in Gribov's approach it is found that $\chi$SB  
takes place if the strong coupling $\alpha_s$ 
exceeds a critical value of $\alpha_c = 0.43$ in some region 
of momenta in the IR. In this situation the dynamical mass function 
$M(q^2)$ exhibits oscillations around zero, and as a consequence the 
relation between the perturbative (or current) quark mass and the 
renormalized mass $m_R=M(0)$ is no longer one-to-one. Instead one 
finds that a nonvanishing renormalized mass can be generated 
even for vanishing perturbative quark mass. It was found though 
that in the approximation discussed above the analytic structure 
of the Green function does not correspond to confined quarks 
\cite{Ewerz:2000qb}. 

When $\chi$SB takes place massless pions are 
created as Goldstone bosons and appear in the physical spectrum. 
Their Bethe-Salpeter amplitude can be obtained as 
$\varphi \sim \left\{ i \gamma_5, G^{-1}\right\}$ from 
an equation for  $q\bar{q}$ bound states derived 
in the same approximation as discussed above for the quark 
\cite{Gribov:1998kb}. 

In the phase of $\chi$SB in which pions exist as (massless) physical 
particles it is natural to consider their backreaction 
on the quark's Green function. In the first approximation described 
above their effect is not properly taken into account. 
Instead, they have to be included explicitly. 
It was argued in \cite{Gribov:1999ui} that pion corrections can 
have a crucial effect in particular on the analytic structure of the 
Green function, possibly giving rise to the confinement of quarks. 

\section{Pion Effects on the Quark's Green Function}

It turns out that the emission and reabsorption of pions can 
be included very easily in Gribov's approximation scheme for 
the quark DSE. 
The pion propagator $\sim 1/k^2$ in connection with 
the differential operator $\partial^2$ makes it again possible to isolate 
the most important IR contribution in diagrams with pion 
loops on the quark. Further, we express the Bethe-Salpeter 
amplitude of the pion in terms of $G$, as a result of which 
we are again left with a differential equation for $G$ only. 
Finally, the pion-quark coupling 
can be fixed with the help of the Goldberger-Treiman 
relation. The resulting equation for $G(q)$ reads
\begin{equation}
\label{gribovgl}
\partial^\mu \partial_\mu G^{-1} = \frac{C_F \alpha_s}{\pi} 
\left(\partial^\mu G^{-1} \right) G \,\partial_\mu G^{-1} 
-\frac{3}{16 \pi^2 f_\pi^2} \left\{ i \gamma_5, G^{-1}\right\}\, G \, 
\left\{i \gamma_5,G^{-1}\right\} \,, 
\end{equation}
where the last term constitutes the modification of the original 
equation due to pions. 

We have performed a numerical study of this improved equation 
for the Green function of light quarks.  
Here we concentrate an the dynamical mass function $M$ in the 
euclidean region, $q^2<0$. More detailed results will be 
presented elsewhere. 

\begin{figure}[ht]
  \includegraphics[height=.29\textheight]{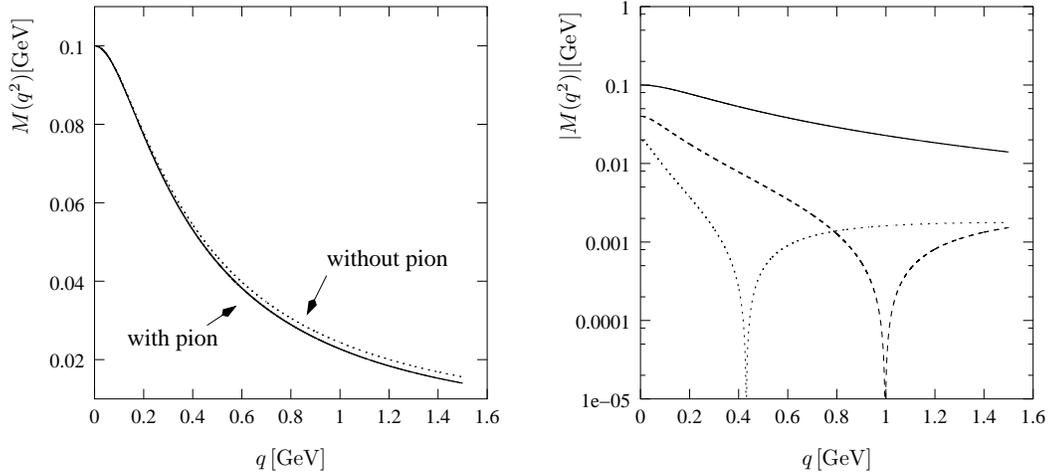}
\caption{Change in the mass function due to pion corrections (left), and the 
mass function for three different values of the renormalized mass (right)
\label{figmass}}
\end{figure}

Recall that the Green function 
can be written as $G^{-1} = Z^{-1} ( {\not \!  q} - M)$ 
with the wave function renormalization factor $Z^{-1}$. 
From equation (\ref{gribovgl}) one finds analytically that the 
pion corrections become negligible in the UV region. Hence 
also the modified equation reproduces the correct RG behavior there. 
Figure \ref{figmass} shows our numerical results for the mass function. 
They have been obtained with an IR frozen coupling $\alpha_s(q)$, 
see \cite{Ewerz:2000qb}. The mass function is affected only very little due 
to the pion corrections (see example on the left). 
The figure on the right shows $M(q^2)$ for different values of 
the renormalized mass. For 
sufficiently small $m_R$ the mass function oscillates, 
giving rise to $\chi$SB. 

In summary, we find only small effects of the pion corrections for 
spacelike momenta. The pattern of chiral symmetry breaking is 
the same as in the approximation without pions. 
Preliminary results indicate that the analytic structure of the 
quark's Green function in the complex $q^2$-plane, on the 
other hand, is considerably changed due to the pion corrections, 
as was anticipated in \cite{Gribov:1999ui}. 
This interesting aspect and its potential consequences for 
confinement clearly deserve further study. 

\begin{theacknowledgments}
This work was supported by a Feodor Lynen fellowship of 
the Alexander von Humboldt Foundation. 
\end{theacknowledgments}


\doingARLO[\bibliographystyle{aipproc}]
          {\ifthenelse{\equal{\AIPcitestyleselect}{num}}
             {\bibliographystyle{arlonum}}
             {\bibliographystyle{arlobib}}
          }
\bibliography{paper}


\end{document}